\documentclass{PoS}

\title{\rule{0cm}{2.5cm}\vspace{-3.5cm}\\{\it \normalsize dedicated to the memory of Thomas Binoth}\\ZZ+jet and Graviton+jet at NLO QCD: recent applications using GOLEM
methods}

\ShortTitle{ZZ+jet and Graviton+jet at NLO QCD: recent applications using
GOLEM methods}

\author{\speaker{Stefan Karg}%
         \thanks{This work is supported in part by the Deutsche Forschungsgemeinschaft
under SFB/TR-9 ``Computergest\"utzte Theoretische Teilchenphysik'',
the Helmholtz Alliance ``Physics at the Terascale'', and the European
Community's Marie-Curie Research Training Network under contract
MRTN-CT-2006-035505 ``Tools and Precision Calculations for Physics
Discoveries at Colliders''}\\
        Institute for Theoretical Physics, RWTH Aachen, D-52056 Aachen, Germany\\
        E-mail: \email{karg@physik.rwth-aachen.de}}

\author{Thomas Binoth\\
     School of Physics and Astronomy, University of Edinburgh, Edinburgh EH9 3JZ, UK %\\
           }
\author{Tanju Gleisberg\\
       SLAC National Accelerator Laboratory, Menlo Park, CA 94025, USA\\
        E-mail: \email{tanju@slac.stanford.edu}}

\author{Nikolas Kauer\\
       Department of Physics, Royal Holloway, University of London, Egham TW20 0EX, UK \\
        E-mail: \email{n.kauer@rhul.ac.uk}}

\author{Gregory Sanguinetti\\
LAPTH, Universite de Savoie, CNRS, BP. 110, 74941 Annecy-le-Vieux, France \\
 E-mail: \email{gregory.sanguinetti@lapp.in2p3.fr}}

\author{Michael Kr\"amer\\
       Institute for Theoretical Physics, RWTH Aachen, D-52056 Aachen, Germany \\
        E-mail: \email{mkraemer@physik.rwth-aachen.de}}

\author{Qiang Li\\
 Paul Scherrer Institut, W\"urenlingen and Villigen, CH-5232 Villigen PSI, Switzerland\\
 E-mail: \email{qiang.li@psi.ch}}

\author{Dieter Zeppenfeld\\
        Institute for Theoretical Physics, Karlsruhe Institute of Technology, D-76128 Karlsruhe, Germany\\
        E-mail: \email{dieter.zeppenfeld@kit.edu}}

\abstract{
In this talk we discuss recent progress concerning precise predictions for hadron colliders. We show results of two applications of tensor reduction using GOLEM methods: the next-to-leading order (NLO) corrections to $pp \to ZZ+$jet  as an important background for Higgs particle and new physics searches at hadron colliders, and the NLO corrections to graviton plus jet hadro-production, which is an important channel for graviton searches at the Tevatron and the LHC. 
}

\FullConference{RADCOR 2009 - 9th International Symposium on Radiative
Corrections (Applications of Quantum Field Theory to Phenomenology) \\
                 October 25-30 2009\\
                 Ascona, Switzerland}

\begin{document}

\section{NLO corrections to $pp \to ZZ+\mathrm{jet}$}
\label{sec:zzj}

Weak boson pair production at hadron colliders plays an essential part in the search for Higgs particles and for physics beyond the Standard Model (SM), since weak bosons can decay into jets, charged leptons or neutrinos and hence produce the same signatures as Higgs bosons, new coloured particles, new electroweak gauge bosons or dark matter candidates.  In addition to being an important background to direct new physics searches at the Large Hadron Collider (LHC) \cite{Campbell:2006wx}, weak boson pair production also allows to search for new physics via experimental evidence for SM deviations in the form of anomalous interactions between electroweak gauge bosons \cite{anomalous_general}. Since LO predictions for hadron collider processes are affected by large QCD scale uncertainties with respect to normalisation and kinematical dependence, the inclusion of NLO QCD corrections is important when comparing predictions for cross sections and differential distributions with data. A process of interest is the production of weak boson pairs with one additional jet at NLO. It is interesting in its own right, due to the enhanced jet activity, particularly at the LHC and in addition provides the real-virtual contribution to the next-to-next-to-leading order (NNLO) corrections to weak boson pair production.  The production of $W$-boson pairs with an additional jet has thus been calculated at NLO without \cite{Dittmaier:2007th} and with \cite{Campbell:2007ev,Dittmaier:2009un} decays. Here, we focus on the process $pp\to ZZj$ which has recently also been computed at NLO \cite{Binoth:2009wk}.

At LO, all channels for $ZZj$ production at hadron colliders are related to the amplitude $0\to ZZq\bar{q}g$ by crossing symmetry.  Therefore, the following subprocesses contribute: \[ 
q\bar{q} \to ZZg\,,\quad
qg \to ZZq\,,\quad
\bar{q}g \to ZZ\bar{q}\,,
\]
where $q$ can be either an up- or down-type quark.
We calculate in the 5-flavour scheme, i.e.~$q=u,c,d,s,b$, and
neglect all quark masses.

\begin{figure}
\centering
\includegraphics[height=4.5cm,angle=0,clip=true]{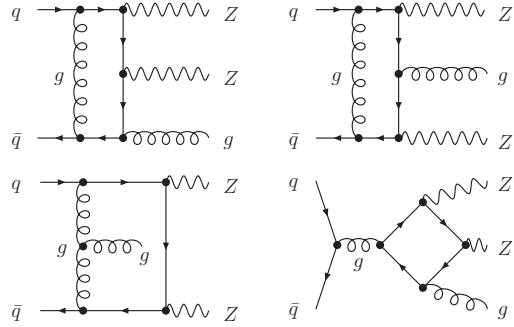}\\[0.2cm]
\caption{%
Representative one-loop graphs for the partonic process $q\bar{q}\to ZZg$.
\label{fig:loop-graphs}}
\end{figure}

At ${\cal O}(\alpha_s)$, the most complicated loop topologies are pentagon graphs derived from the tree-level graphs via virtual gluon exchange (and crossing), and box graphs derived by closing the quark line in the tree-level graphs and attaching a $gq\bar{q}$ current. Representative one-loop graphs for the partonic process $q\bar{q}\to ZZg$ are shown in Fig.~\ref{fig:loop-graphs}.
Two independent sets of amplitude expressions have been generated, both of them using the spinor helicity formalism of Ref.~\cite{Xu:1986xb}.  Polarisation vectors have been represented via spinor traces, i.e.~kinematic invariants up to global phases.  By obtaining an analytical representation for the full amplitude, we aim at promoting simplification via analytical cancellations. Especially we employ that, apart from the rank one case, all pentagon tensor integrals are reducible, i.e.~can directly be written as simple combinations of box tensor integrals.   For the remaining tensor integrals we employ the GOLEM-approach \cite{golem_reduction}.

The ${\cal O}(\alpha_s)$ real correction channels for $ZZj$ production at hadron colliders are related to the amplitudes $0\to ZZq\bar{q}gg$ and $0\to ZZq\bar{q}q'\bar{q}'$ by crossing symmetry.  While all virtual correction channels are already present at LO, new real correction channels open up at NLO, namely the $gg$, $qq'$, $q\bar{q}'$ ($q'\neq q$) and $\bar{q}\bar{q}'$ channels. Note that these new channels are effectively of LO type. To facilitate the cancellation of soft and collinear singularities we employ the Catani-Seymour dipole subtraction method \cite{Catani:1996vz}. We use the SHERPA implementation \cite{Sherpa} to calculate numerical results for the finite real correction contribution.

\begin{figure}
\centering
\begin{minipage}[c]{.49\linewidth}
\flushleft \includegraphics[height=7.3cm,angle=0,clip=true]{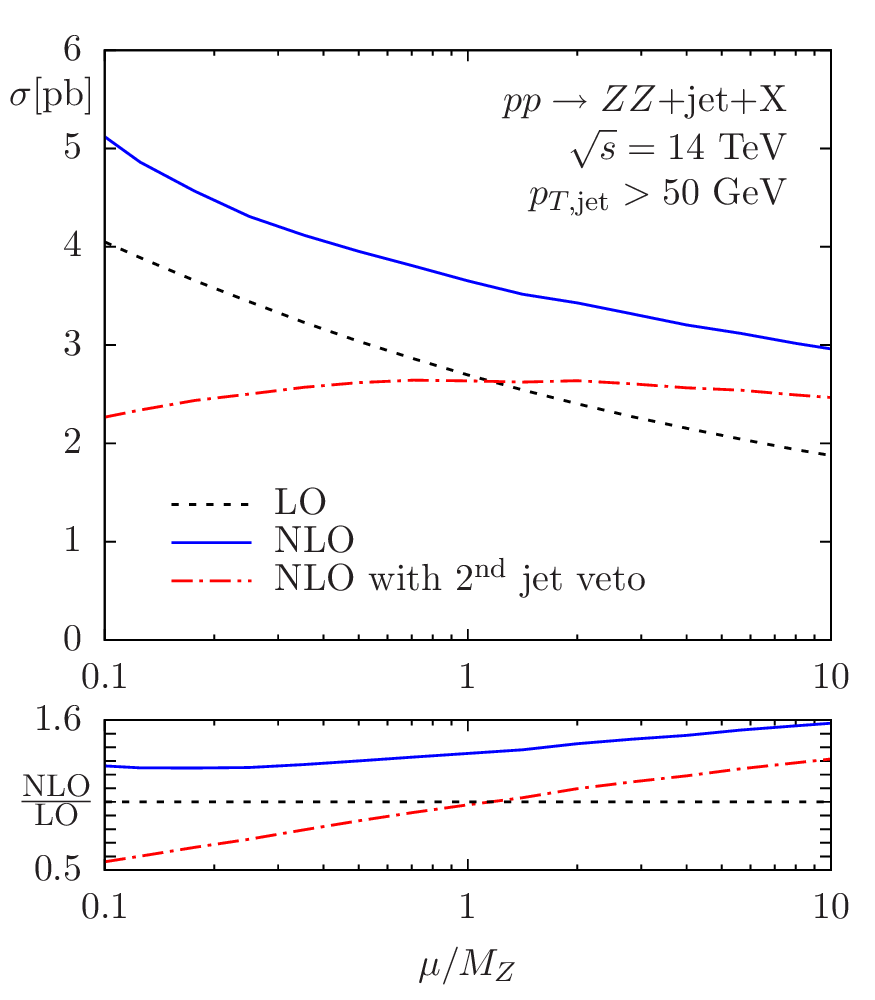}
\end{minipage} \hfill
\begin{minipage}[c]{.49\linewidth}
\flushright \includegraphics[height=7.3cm,angle=0,clip=true]{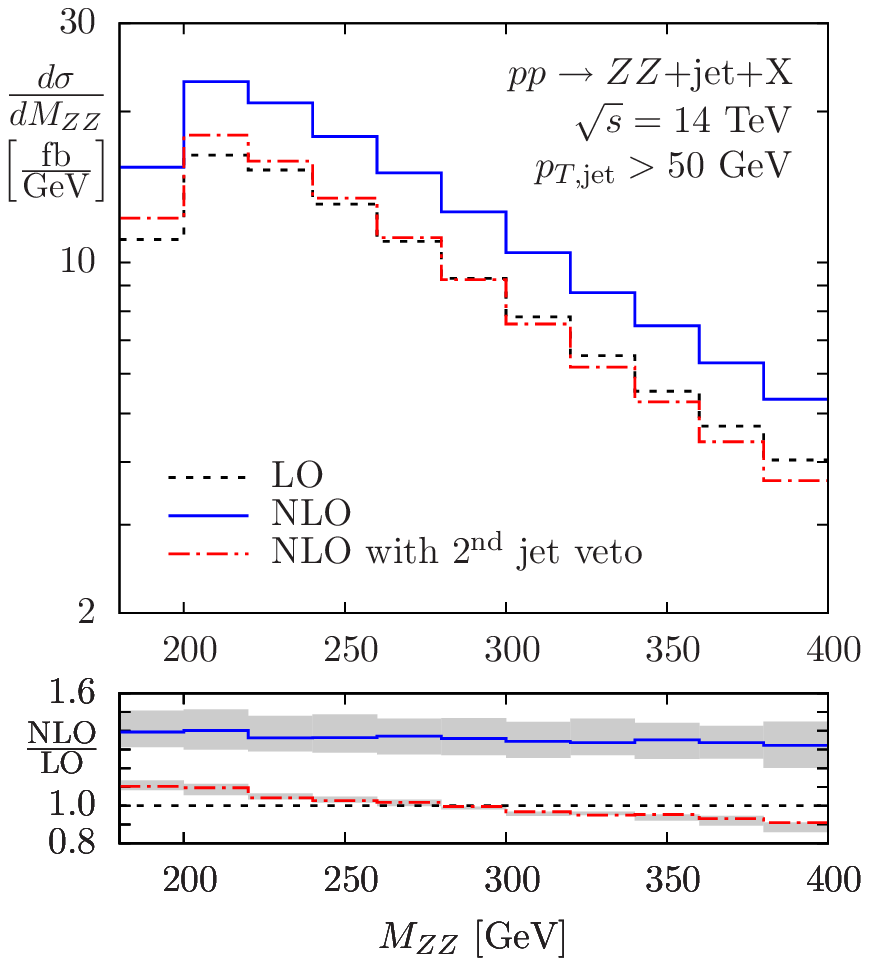}
\end{minipage}%\\[0.2cm]
\caption{%
Scale dependence ($\mu_R=\mu_F=\mu$) of the $ZZ$+jet cross section and  the $ZZ$ invariant mass distribution   
at the LHC with $p_{T,\,\textrm{jet}} > 50$ GeV for the hardest jet 
in LO (dotted) and NLO (solid).  The exclusive NLO cross section when a
$p_{T,\textrm{jet}} > 50$ GeV veto for additional jets is applied is also
shown (dot-dashed).
\label{fig:scalevar}}
\end{figure}

In Fig.~\ref{fig:scalevar}, LO and NLO predictions for $ZZj$ production 
cross sections at the  LHC are displayed.
The shape of the cross section scale variation at the LHC  is qualitatively unchanged when going from LO to NLO, in contrast to the Tevatron, where the cross section reaches its maximum at approximately $M_Z/2$ and where its variation is very effectively reduced. We attribute this to new channels that become active at NLO. These channels have a modest impact at the Tevatron, but a sizable impact at the LHC, due to parton densities being probed in different $x$ regions.  We also calculate an exclusive NLO cross section for the LHC by vetoing 2-jet events with a second hardest jet with $p_T > 50$ GeV (NLO with 2$^\textrm{\scriptsize nd}$ jet veto). This exclusive NLO LHC cross section decreases for scales below $M_Z$ and has a strongly reduced scale uncertainty.  In general, the $K$ factor for $ZZj$ production will have a non-negligible dependence on the kinematics.  As an example, we display in Fig.~\ref{fig:scalevar} the differential LO and NLO distributions with respect to the invariant $ZZ$ mass and the resulting $K$ factor at the  LHC. The $K$-factor bands shown in this figure correspond to a variation of the scale $\mu$ by a factor of $2$ in the NLO differential cross section only, i.e.~we display $[\rm{d}\sigma_\textrm{NLO}/dM_{ZZ}](\mu)/[\rm{d}\sigma_\textrm{LO}/\rm{d}M_{ZZ}](M_Z)$
with $\mu/M_Z\in[\frac{1}{2}, 2]$.

\section{NLO corrections to $pp \to G+\mathrm{jet}$}
\label{sec:Gj}

The search for new physics at the TeV-scale is one of the major tasks
for current and future high-energy physics experiments.  Models with
extra space dimensions and TeV-scale gravity address the problem of
the large hierarchy between the electroweak and Planck scales, and
predict exciting signatures of new physics that can be probed at
colliders~\cite{Giudice:2008zza}.

In the $D=4+\delta$ dimensional model proposed by Arkani-Hamed,
Dimopoulos and Dvali (ADD)~\cite{ADD}, the SM
particles are constrained to a $3+1$ dimensional brane, while gravity
can propagate in a $4+\delta$ dimensional space-time.  For simplicity,
the additional $\delta$-dimensional space is assumed to be a torus
with common compactification radius $R$. In such a model, the
4-dimensional effective Planck scale $M_{\rm P}$ is related to the
fundamental scale $M_{\rm S}$ by~\cite{ADD}:
\begin{equation}\label{scale}
M^2_{\rm P}=8\pi R^\delta M^{\delta+2}_{\rm S}\,.
\end{equation}
For a large compactification radius $R$ it is thus possible that the
fundamental scale is near the weak scale, $M_{\rm S} \sim {\rm TeV}$.

The $D=4+\delta$ dimensional graviton corresponds to a tower of massive 
Kaluza-Klein (KK) modes in 4 dimensions. Although each individual graviton 
couples to SM matter with only gravitational strength $\propto 1/{ M}_{\rm P}$,  
inclusive collider processes, where one sums over all accessible KK modes, 
are enhanced by their enormous number $\propto { M}_{\rm P}^2$ leaving an 
overall suppression of only $M_{\rm S}^{-2-\delta}$. If the fundamental 
scale $M_{\rm S}$ is near the TeV-scale, graviton production can thus be 
probed at present and future high-energy colliders.

Both virtual graviton exchange between SM particles and real graviton
emission provide viable signatures of large extra dimensions at
colliders. Since the coupling of gravitons with matter is suppressed
$\propto 1/{ M}_{\rm P}$, direct graviton production gives
rise to missing energy signals. Searches for graviton production have
been performed in the processes $e^+e^- \to \gamma(Z) + E^{\rm miss}$
at LEP and $p\bar{p}\to \gamma({\rm jet}) + p_T^{\rm miss}$ at the
Tevatron~\cite{LEP}. Searches for the
process $pp\to {\rm jet} + p_T^{\rm miss}$ at the LHC will be able to
extend the sensitivity to the fundamental scale $M_{\rm S}$ into the
multi-TeV region~\cite{LHCG, Wu:2008hw}.

The NLO QCD corrections to graviton production in the process $pp/p\bar{p}\to {\rm
  jet} + G$ have been computed recently \cite{Karg:2009xk}.
The NLO cross sections lead to significantly more accurate theoretical signal 
predictions and thereby more accurate constraints on $M_{\rm S}$ or, in the case 
of discovery, will allow to probe the model parameters. 

The LO cross section for graviton plus jet production receives
contributions from the partonic processes
\begin{equation}\label{eq:xshat}
  q\bar{q}\rightarrow gG,\quad qg\rightarrow qG \quad {\rm and}\quad
  gg\rightarrow gG\,.
\end{equation}
We have performed two independent calculations of the virtual corrections: the first calculation is based on the Mathematica package {FeynCalc}~\cite{FeynCalc}. Because of the Lorentz indices of the spin-2 graviton, we encounter high-rank tensor integrals, such as rank-5 4-point functions. Special care is taken to reduce those to one-loop scalar integrals by an independent Mathematica code, following the prescription of Ref.~\cite{rank5}.

The second calculation is based on the GOLEM-approach~\cite{golem_reduction} as described in section~\ref{sec:zzj}.
%The setup of the second calculation closely follows the one described  in section~\ref{sec:zzj}. 
Only tensor reduction routines for rank $N+1$ $N$-point tensor integrals with $N\leq 3$ had to be added.

We have checked gauge invariance and Ward identities arising from general coordinate invariance, see Ref.~\cite{G2j} for more details. The numerical implementation of the real-emission contributions is based on { MadGraph}~\cite{spin2MG} and {MadDipole}~\cite{MadDipole}. %

\begin{figure}
\centering
\begin{minipage}[c]{.49\linewidth}
\flushleft \includegraphics[height=5.0cm,angle=0,clip=true]{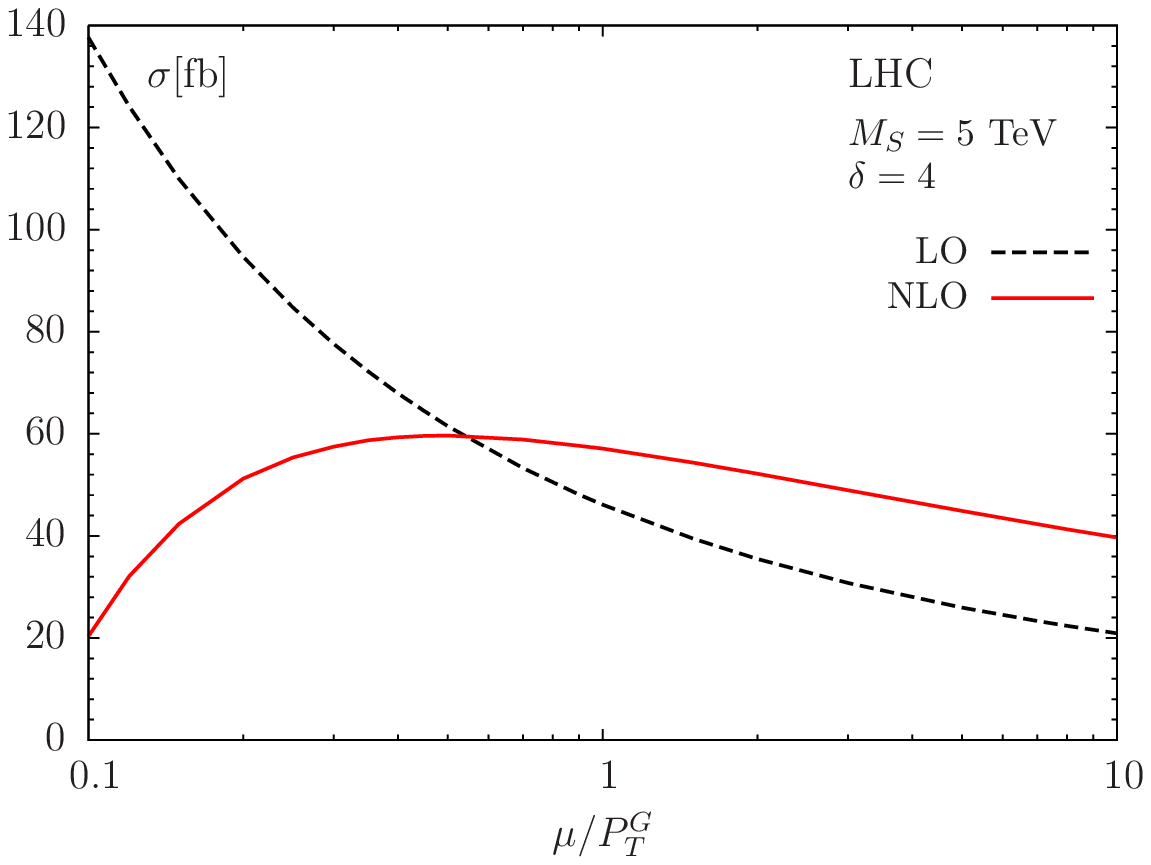}
\end{minipage} \hfill
\begin{minipage}[c]{.49\linewidth}
\flushright \includegraphics[height=5.0cm,angle=0,clip=true]{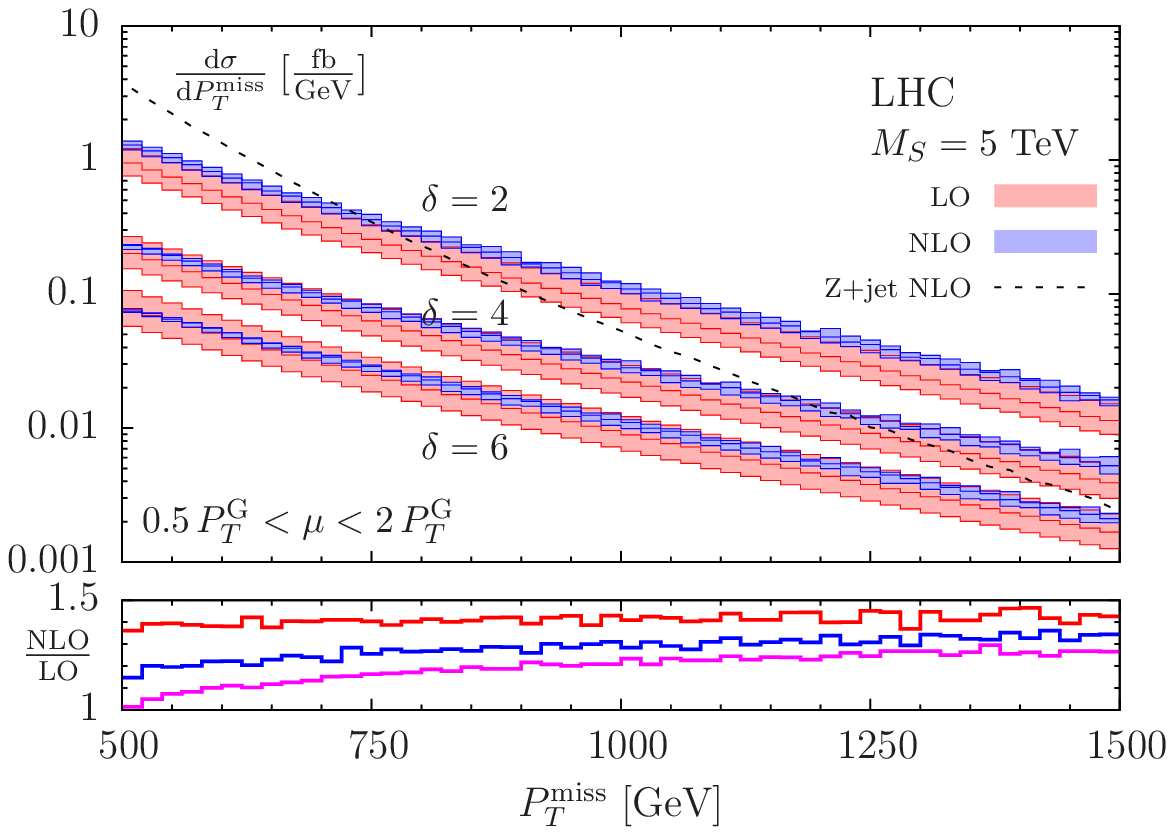}
\end{minipage}\\%[0.2cm]
\caption{%
Left: scale variation for the integrated cross section at LHC and
Tevatron, for a common scale $\mu=\mu_r=\mu_f$ and $P_T^{\mathrm{miss}}> 500$~GeV.
Right: $P_T^{\mathrm{miss}}$ distribution for the graviton signal at
the LHC with scale uncertainty bands ($0.5 \,P_T^{\mathrm{G}} < \mu
<2\,P_T^{\mathrm{G}}$). Also given is the NLO distribution for the
dominant $Z\to \nu\bar\nu$ background. The lower part of the plot
shows $K(P_T)=(\rm{d}\sigma_{\rm NLO}/\rm{d}P_T)/(\rm{d}\sigma_{\rm LO}/\rm{d}P_T)$ for
$\delta=2,4,6$ (top down).
\label{fig:Gjplots}}
\end{figure}

In Fig.~\ref{fig:Gjplots}, LO and NLO predictions for Graviton plus jet production 
cross sections at the LHC are displayed.
We observe that the scale dependence of the NLO cross section is significantly smaller than that of the LO cross section: changing $\mu$ in the range between $P_T^G/2 $ and $2P_T^G $, the LO cross section varies by $\approx 30\%$, while the scale uncertainty at NLO is less than $\approx 10\%$. At the LHC, the $K$-factor, $K=\sigma_{\rm NLO}/\sigma_{\rm LO}$, is sizeable at the central scale $\mu = P_T^G$, increasing the LO cross section prediction by about $20\%$.

The experimental analyses at the LHC  rely on the $P_T^{\mathrm{miss}}$  distribution. The right plot in Fig.~\ref{fig:Gjplots} shows the scale dependence of this distribution, for different choices of the number of extra dimensions $\delta=2,4,6$.  We also show the NLO QCD predictions for the main background $pp\rightarrow Z(\to\nu\bar{\nu})+$\,jet obtained with {\ MCFM}~\cite{mcfm}. The bands show the uncertainty of the LO and NLO predictions when varying the renormalization and factorization scales in the range $P_T^G/2 < \mu <2\,P_T^G$. The reduction of the scale uncertainty at NLO is evident. The figure  also displays the $P_T$ dependence of the $K$ factors, defined as $K(P_T)=(\rm{d}\sigma_{\rm NLO}/\rm{d}P_T)/(\rm{d}\sigma_{\rm LO}/\rm{d}P_T)$. The NLO corrections  are sizeable at the LHC  and increasing with decreasing $\delta$.  Furthermore, the $K$ factors depend on the kinematics and increase with increasing $P_T^{\mathrm{miss}}$.

\section{Dedication}
We dedicate  this proceedings contribution to our colleague and dear friend Thomas Binoth. We honour him as a great and passionate physicist, and will remember him as a warmhearted, honest and wonderful  friend  who will be greatly missed.

\end{document}